# Controlled lasing from active optomechanical resonators

T. Czerniuk[1]*, C. Brüggemann[1], J. Tepper[1],  S. Brodbeck[2], C. Schneider[2], M. Kamp[2], S. Höfling[3], B. A. Glavin[4], D. R. Yakovlev[1,5], A. V. Akimov[5,6], and M. Bayer[1]

[1]*Experimentelle Physik 2, TU  Dortmund University, 44227 Dortmund, Germany*

[2]*Technische Physik, Physikalisches Institut and Wilhelm Conrad Röntgen-Center for Complex Material Systems, Universiy of Würzburg, Am Hubland, 97074 Würzburg, Germany*

[3]*School of Physics and Astronomy, University of St Andrews, St Andrews KY16 9SS, United Kingdom*

[4]*V. E. Lashkaryov Institute of Semiconductor Physics, 03028 Kyiv, Ukraine*

[5]*A. F. Ioffe Physical-Technical Institute, Russian Academy of Sciences, 194021 St. Petersburg, Russia*

[6]*School of Physics and Astronomy, University of Nottingham, Nottingham NG7 2RD, United Kingdom*

* Thomas.Czerniuk@tu-dortmund.de

**Planar microcavities with distributed Bragg reflectors (DBRs) host, besides confined optical modes, also mechanical resonances due to stop bands in the phonon dispersion relation of the DBRs[1,2]. These resonances have frequencies in the sub-terahertz ($10^{10}$-$10^{11}$ Hz) range with quality factors exceeding 1000. The interaction of photons and phonons in such optomechanical systems can be drastically enhanced, opening a new route toward manipulation of light[3,4]. Here we implemented active semiconducting layers into the microcavity to obtain a vertical-cavity surface-emitting laser (VCSEL). Thereby three resonant excitations - photons, phonons, and electrons – can interact strongly with each other providing control of the VCSEL laser emission: a picosecond strain pulse injected into the VCSEL excites long-living mechanical resonances therein. As a result, modulation of the lasing intensity at frequencies up to 40 GHz is observed. From these findings prospective applications such as THz laser control and stimulated phonon emission may emerge.**

Optomechanical devices are structures that host two different types of resonances: one resonance for electromagnetic waves (photons) and another one for high-frequency mechanical vibrations (phonons). The wealth of predicted novel phenomena in these systems has attracted huge interest to them[5-10]. Optomechanical structures offer great flexibility of possible implementations ranging from single free-standing nanostructures (e.g., spheres, toroids, cantilevers and membranes)[11] to periodic arrays of these systems[4,12,13]. The size of nanostructures and the period in arrays determine the resonance frequencies of photons and phonons. For near-infrared or visible light these dimensions have to be in the sub-micrometer range. The concomitant resonances for the phonons have frequencies in the sub-THz ($10^{10}$-$10^{11}$ Hz) range so that they could be excited and operated by picosecond acoustic techniques[14].

All experiments performed so far on mechanically driven optomechanical devices have involved optically-passive systems[3-13]. A major step forward in implementing such devices into integrated optoelectronic circuits would be the extension towards active structures that can generate light. A device particularly important for telecommunication applications is the vertical-cavity surface-emitting laser (VCSEL)[15]. VCSELs contain optically active media [e.g. semiconductor quantum wells (QWs) or quantum dots (QDs)] embedded in a planar microcavity (MC). They are excited optically or electrically in order to reach population inversion between the two electronic states involved in stimulated emission. Ideally the energy separation between these states matches the optical cavity mode energy.

Mechanical waves in a MC with two distributed Bragg reflectors (DBRs) may be considered in a way similar to confined electromagnetic waves. Due to the acoustic impedance mismatch in the DBR layers, stop-bands emerge due to frequency splittings in the phonon dispersion at the center and at the edges of the folded Brillouin zone scheme[1,2,4]. Two types of mechanical resonances may exist in such resonators: strongly localized MC phonon modes within the DBR stop bands that arise due to the imperfection of the DBR periodicity introduced by the central cavity layers; and DBR resonances with a large density of states at the edges of the stop-bands. Following the acousto-optical studies by *Fainstein et al* on a passive MC[4], ideas of exploring and exploiting the optomechanical properties of VCSELs have been around for some time[16] but so far no corresponding experimental or theoretical investigation with VCSELs in active lasing operation have been carried out.

Particularly challenging in studying the impact of optomechanical VCSEL resonances is the understanding of the phonon interaction processes with the optically active medium. Many different relevant processes have to be taken into account, and in general the lasing dynamics are nonlinear and may involve competing resonator modes[17]. To develop such understanding requires a large scale effort. However, before starting such an extensive effort, its value should be evidenced by demonstrating that the dual resonances in active optomechanical devices like VCSELs do indeed have a significant impact on the lasing output - the present work answers this question favorably.

We report picosecond acoustic experiments carried out on two VCSELs, for which we have chosen on purpose very different designs to highlight the potential of our approach. The main result is observation of long-living and high-amplitude harmonic oscillations in the lasing output due to excitation of mechanical VCSEL resonances. The experiment is shown schematically in Fig. 1a. Light emission from the optically active media in the VCSELs is result of continuous or pulsed laser excitation while the coherent phonons are excited remotely by injection of strain pulses from the VCSEL substrate. The strong impact of resonant phonons is evidenced by a pronounced ringing of the laser emission, which continues for nanoseconds long after the original broadband picosecond strain pulse has left the active region of the VCSEL.

The two VCSELs were based on GaAs/AlAs material, with the first one (VCSEL1) containing 12 QW layers distributed over three stacks and the second one (VCSEL2) containing a single QD layer only. These optically active layers were placed in the electric field antinodes of the central cavity layer or the adjacent layers, sandwiched between the DBRs, as shown in Fig. 1a. Figure 1b shows calculations of the phonon dispersion relations of the DBRs in these two samples, with the close-ups highlighting the band gap regions. At these band gaps, the phonon group velocity tends to zero, corresponding to particularly long-living phonons inside the cavities. The coherent phonon wavepacket in form of a broadband picosecond strain pulse $\varepsilon_0(z - st)$ with $s$=4.8×10³ m/s being the longitudinal sound velocity in GaAs, is generated in a 100 nm thick Al film deposited on the GaAs substrate opposite to the cavity[18], as shown schematically in Fig. 1a. Following injection into the VCSEL, the strain pulse propagates sequentially through bottom DBR, MC layer, and finally top DBR. After reflection at the surface the strain pulse passes again through these layers in reversed order. During propagation it undergoes spatio-temporal transformation due to the multiple reflections at the interfaces of the multilayered structure. The temporal strain profiles, calculated by the transfer matrix formalism at the position of the central cavity layer in the two VCSELs, are shown in Fig. 1c. In the time window when the strain pulse is outside, still a nanosecond tail is present in this layer due to the mechanical VCSEL resonances that have been excited by the acoustic pulse. The corresponding Fourier spectra are shown in the inset of Fig. 1c, demonstrating clearly the excitation of resonant modes with frequencies coinciding with the band edges in the phonon dispersion. For our particular cases the peaks correspond to DBR resonances at the edges of the first and second band gaps at the zone-border and zone-center, respectively, while MC resonances do not show up [see Supplementary Information 1].

Besides their different composition, the VCSELs were operated at different temperatures applying varying optical excitation conditions, which further highlights the versatility of our method. VCSEL1 was excited by femtosecond pulses and the resulting emission intensity $I(t)$ was measured time-integrated by a photodiode as function of delay $\tau$ between optical excitation of laser emission and picosecond strain pulse injection. We present here data recorded for excitation far above the laser threshold. Figure 2 shows the measured signal $\Delta I(\tau)/I_0$ ($I_0$ is the signal in absence of the strain pulse) in the delay time interval when the strain pulse is travelling through the VCSEL. Panels a and b show the measurements at $T$=300 K, room temperature, and at $T$=180 K, respectively. By this temperature change the quantum well emission is considerably shifted relative to the cavity mode, as seen from the lower right-hand insets in Fig. 2. These show the spontaneous emission spectra recorded from the edge of VCSEL1.

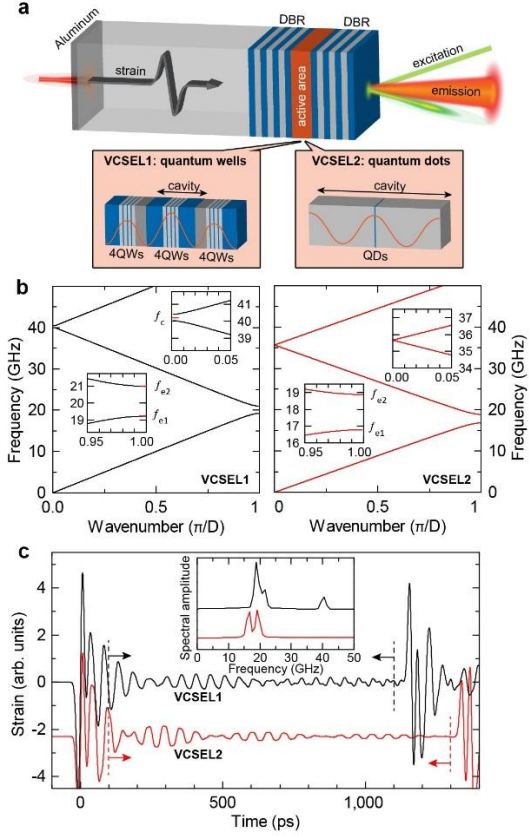

**Figure 1: Experimental technique for studying mechanical resonances in VCSELs**. a, Scheme of the experiment showing the strain pulse propagating through the VCSEL. The zooms into the VCSEL structure give the optical field intensity distributions in the two used cavities. The picosecond strain pulses are excited by optical excitation of the Al film deposited on the substrate side opposite to the VCSEL. Pulsed or cw-optical excitation is used to push the VCSELs into lasing. b, Dispersion relations for LA phonons in the DBRs of the two different VCSELs. The close-ups show the parts of the folded Brillouin zone scheme, where band gaps are observed so that mechanical resonances, which are long-living in the VCSELs due to their very small group velocities, may be excited. VCSEL2 does not possess a stop band in the zone-center and correspondingly no resonance around 36 GHz exists. c, Calculated temporal evolutions of the picosecond strain pulse in the middle of the VCSEL's cavities. The two temporal intervals with high-amplitude oscillations correspond to the incident and reflected broadband strain pulses being inside the cavity while low amplitude oscillations in between are due to long-living resonance phonon modes of the VCSELs. Inset: Spectra of the strain pulses obtained by Fourier transforming their temporal evolutions in the time windows marked by the arrows and the vertical dashed lines.

The corresponding temporal evolutions $I(t)$ of the laser output pulses are given in the lower left-hand insets.

The intensity modulations $\Delta I(\tau)/I_0$ in Figs. 2a and 2b show prominent oscillations for whose assignment several time ranges, labeled A-E, are indicated with their borders shown by the vertical dash-dotted lines. These time ranges are defined by the momentary position of the vraodband picosecond strain pulse. Maximum amplitude of the oscillations occurs in ranges B and D when the leading edges of the strain pulses pass the QWs in forward and backward directions, respectively, as highlighted also by the horizontal bars in Fig. 2b. Smaller amplitude oscillations of $\Delta I(\tau)/I_0$ are observed in the C and E ranges when the tail of the coherent phonon pulse initiated by the DBR is present in the cavity layer with the QWs. The Fourier spectra of $\Delta I(\tau)/I_0$ in these ranges in Figs. 2c and 2d exhibit several distinct resonance peaks. Most strikingly these resonance coincide with the calculated frequencies for the zone-edge ($f_{e1}$, $f_{e2}$) and zone-center ($f_c$) resonant phonons with long cavity lifetimes due to their small group velocity in VCSEL1 (compare Figs. 2c,d with the inset in Fig. 1c). This excellent agreement between measured and calculated frequencies demonstrates that the laser emission is modulated by the resonant phonons of the active optomechanical device.

Our interpretation is corroborated by the results obtained for VCSEL2, which was optically excited by a continuous wave laser with a wavelength of 532 nm. Here, the impact of the picosecond strain pulses was monitored in real time by measuring $I(t)$ with a streak camera, again well above the lasing threshold. For this sample, a temperature of $T$=10 K was chosen to have a significant detuning between the QD emission with maximum at $E_{QD}$=1.351 eV and the cavity mode at $E_{c2}$=1.367 eV, which still is sufficient to obtain lasing (inset in Fig. 3a). The measured temporal evolution $\Delta I(\tau)/I_0$ of the modulation enforced by the incident strain pulse is shown in Fig. 3a. The modulation peak around time zero does not exploit the phonon band structure of the DBRs, but can be understood by considering the VCSEL as an elastic continuum as discussed in earlier work[19]. The striking novel feature of interest here is the pronounced oscillating tail in $I(t)$ extended over 1.5 ns. The Fourier spectrum of this tail is shown in Fig. 3b and consists of the resonant peak at $f_{e1}$=16.9 GHz, corresponding again exactly to the lower energy DBR zone-edge phonon resonance of VCSEL2 (compare with calculated spectrum in Fig. 1c). Thus, similar to VCSEL1, the lasing emission from VCSEL2 shows clear fingerprints of optomechanical resonances. The absence of the peak corresponding to the zone-center in VCSEL2 is due to perfect acoustic matching in the DBR layers which results in vanishing of the phonon stop-band at the Brillouin zone center (see Fig. 1b).

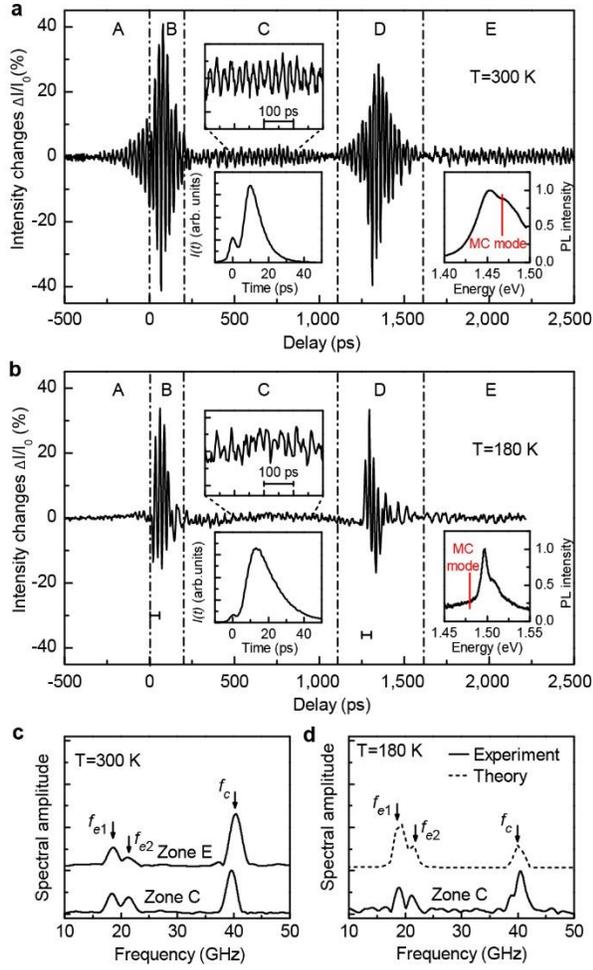

**Figure 2: Modulation of the lasing output in VCSEL1. a** and **b,** Time integrated lasing output as function of the delay between laser cavity and strain pulse excitations measured for VCSEL1 at $T$=300 K (**a**) and 180 K (**b**). The vertical dash-dotted lines indicate the borders between the temporal zones A-E. Zones C and E correspond to the ranges in which long-living resonant phonons of the active optomechanical device persist in the cavity layer containing the optically active medium. Close-ups of the temporal evolutions of the oscillating signals are shown in the upper insets in zone C. The lower insets in zone C show the time evolutions of the laser pulses; the small peak at the beginning corresponds to the optical excitation of the VCSEL1. The right insets in zone E show the spectral position of the MC optical modes relative to the PL spectrum, measured from the side of the VCSEL. **c** and **d,** Fourier spectra obtained from the temporal evolutions in zones C and E shown in **a** and **b,** respectively. The spectra show peaks which are in excellent agreement with the resonant frequencies of the zone-edge ($f_{e1}$ and $f_{e2}$) and zone-centre ($f_c$) phonons in VCSEL1. The dotted curve in **d** is the calculated spectrum assuming the deformation potential mechanism for the strain-induced modulation of the lasing output from the 12 active QWs in VCSEL1.

Concerning the mechanisms responsible for modulating the laser emission by coherent phonons we consider two major contributions: (1) modulation of the gain spectrum of QWs and QDs in VCSEL1 and VCSEL2, respectively; and (2) switching between different optical laser modes in VCSEL1. The first one is governed by the strain ($\varepsilon$) induced energy shift of the band gap $\delta E_{QW(QD)}$=$\Xi\varepsilon$, where $\Xi\sim 10$ eV is the sum deformation potential of conduction and valence bands. This mechanism gives reasonable agreement between simulated and measured $\Delta I(\tau)/I_0$ in VCSEL1 for high pumping densities when the onset of laser emission is faster than the period of mechanical resonances. The calculated Fourier spectrum of the oscillating tail is shown in Fig. 2d by the dotted curve [see Supplementary Information 2]. Reasonable agreement is also obtained for VCSEL2, for which the inhomogeneous distribution of exciton resonances in the QDs[20] was included in the analysis of the deformation potential mechanism. The calculated amplitude spectrum is shown in Fig. 3b by the dashed line along with the experimental Fourier transform as solid line.

However, the deformation potential model cannot explain the modulation occurring for VCSEL1 in time range A before the strain pulse reaches the QWs, most prominent at 300 K, see Fig. 2a. In this interval the strain pulse is traveling through the first DBR so that it cannot modulate the spectral profile of the QW gain medium. Here most likely the phonons modulate the optical mode distribution of the laser emission[17].

In conclusion our experiments evidence the interplay of the optical and acoustical dual resonance properties of VCSEL devices operated in the lasing regime, resulting in an emission modulation by resonant coherent phonons at 40 GHz frequency, even at room temperature. The modulation amplitude reaches significant values: up to 4% in VCSEL1 and 50% in VCSEL2. The wealth of underlying mechanisms (deformation potential coupling, optical mode switching etc) offer a variety of tools for controlling the light emission, depending on the specific design of the active optomechanical device.

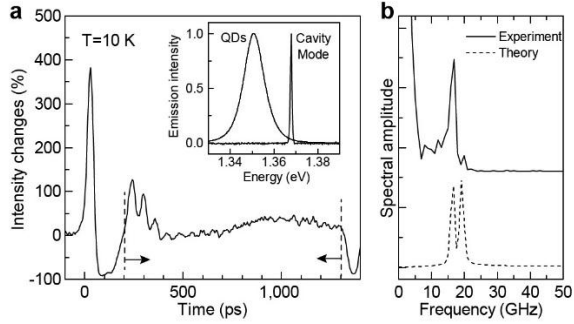

**Figure 3: Modulation of the lasing output in VCSEL2. a,** Time evolution of the lasing output due to cw-excitation of the QDs in the cavity, recorded over the time interval when the picosecond strain pulse is passing VCSEL2, measured at $T=10$ K. Of particular interest are the oscillations at $t>200$ ps, whose frequency spectrum obtained by Fourier transforming the time-resolved emission in the time window marked by the arrows and the dashed lines is shown in **b.** The spectrum exhibits a pronounced peak at a frequency of 16.9 GHz, which agrees well with the peak corresponding to the zone-edge resonance phonons of VCSEL2. The inset in **a** shows the PL emission spectrum of the QDs recorded from the side of the cavity relative to the optical cavity mode.

We believe that our study marks the starting point for using optomechanical properties of VCSELs and other laser resonator structures to modulate their emission in nanophotonic circuits. Beyond exploiting the optical resonator DBRs, also design of planar optomechanical devices with separate photon and phonon DBRs is feasible by which the acoustic resonance frequencies could be raised into the THz range[21]. Thus, in prospective, optical clocking on THz frequencies may become feasible, which is not possible by traditional methods. Active optomechanical devices also open a novel approach for realizing coherent stimulated phonon emission (phonon laser or in short "saser"[22,23]). In comparison with passive devices, the interaction between photons and phonons is increased by several orders of magnitude due to electron-phonon interaction in the active medium. The high sensitivity of the laser structure to dynamical strain in VCSELs resembles a tool for obtaining phonon lasing and exploiting opto-acoustic phenomena in integrated active optomechanical devices.

**Methods**

**Samples and lasing parameters.** Both devices were grown by molecular-beam epitaxy on n-doped [001] oriented GaAs substrates. Three groups of 12 nm wide GaAs QWs form the active region of VCSEL1. Each group consists of 4 QWs separated from each other by 4 nm wide AlAs barriers. The middle group is located in the center of the $\lambda/2$-cavity, while the outer QW sets are placed in the first DBR layers, where the photon field has an antinode, as shown in Fig. 1a. The DBRs are built out of 58 nm wide $Al_{0.2}Ga_{0.8}As$ and 67 nm wide AlAs double-layers corresponding to $\lambda/4$ optical thickness (27 and 23 periods in the bottom and top reflector, respectively). The Q-factor $\sim 10^4$ of the optical cavity was deduced from the calculated reflectivity spectrum and confirmed by local micro-photoluminescence studies at 10 K. In order to obtain the spectrum of the gain medium unfiltered by the MC, the QW photoluminescence was collected from the side of the sample. For excitation of VCSEL1 laser pulses from a Ti:sapphire oscillator were used (800 nm center wavelength, 140 fs pulse duration and 100 kHz repetition rate). At room temperature, the lasing threshold is crossed at 600 μJ/cm². At $T=180$ K, the threshold increases significantly to 1,600 μJ/cm², due to the narrowing of the spectral profile of the electron-hole transition and its high-energy shift relative to the cavity mode (see the right inset in Fig. 2b). The angular aperture of the detection was about 2°, smaller than the beam divergence of common VCSELs[24].

The DBR stacks sandwiching the GaAs $\lambda$ cavity layer of VCSEL2 are composed of alternating 67 nm wide GaAs and 80 nm wide AlAs double-layers (27 pairs at the bottom, 23 pairs at the top). A single sheet of $In_{0.3}Ga_{0.7}As$ QDs in the center of the cavity layer acts as active region, see Fig. 1a. The ensemble is inhomogeneously broadened with a spectral width of 11 meV at $T=10$ K, much more than the cavity mode linewidth of 1.2 meV, which results in an inefficient coupling. VCSEL2 is optically excited using a Q-switched Nd:YAG laser emitting 23 ns pulses at 532 nm. The excitation pulse duration is much longer than the laser emission dynamics, so that the lasing intensity follows adiabatically and may be treated as stationary. At a peak excitation power density of 22 kW/cm², the lasing threshold is crossed.

**Generation of picosecond strain pulses.** A 100-nm-thick Aluminum film was deposited on the GaAs substrate opposite to the microcavity. For generation of the strain pulses, it is illuminated by laser pulses from the same Ti-sapphire laser used for pushing VCSEL1 into lasing. The energy density per pulse is $\sim 10$ mJ/cm², resulting in injection of a strain pulse (schematically shown in Fig.1a) with amplitude of $\sim 10^{-3}$ and duration of $\sim 10$ ps[18].